\documentclass[preprint2]{aastex}
\slugcomment{Resubmitted to PASP, October 2007.}

\shorttitle{Possible Associations between PNe and Open Clusters}
\shortauthors{Majaess, Turner \& Lane}

\begin{document}

\title{In Search of Possible Associations between Planetary Nebulae and 
    Open Clusters}

\author{Daniel J. Majaess and David G. Turner}
\affil{Department of Astronomy and Physics, Saint Mary's University,
    Halifax, Nova Scotia B3H 3C3, Canada}
\email{dmajaess@ap.smu.ca, turner@ap.smu.ca}

\and

\author{David J. Lane}
\affil{The Abbey Ridge Observatory, Stillwater Lake, Nova Scotia, Canada}
\email{dlane@ap.smu.ca}

\begin{abstract}
We consider the possibility of cluster membership for 13 planetary nebulae 
that are located in close proximity to open clusters lying in their lines 
of sight. The short lifetimes and low sample size of intermediate-mass 
planetary nebulae with respect to nearby open clusters conspire to reduce 
the probability of observing a true association. Not surprisingly, line of 
sight coincidences almost certainly exist for 7 of the 13 cases considered. 
Additional studies are advocated, however, for 6 planetary nebula/open 
cluster coincidences in which a physical association is not excluded by the 
available evidence, namely M 1-80/Berkeley 57, NGC 2438/NGC 2437, 
NGC 2452/NGC 2453, VBRC 2 \& NGC 2899/IC 2488, and HeFa 1/NGC 6067. A number 
of additional potential associations between planetary nebulae and open 
clusters is tabulated for reference purposes. It is noteworthy that the 
strongest cases involve planetary nebulae lying in cluster coronae, a 
feature also found for short-period cluster Cepheids, which are themselves 
potential progenitors of planetary nebulae. 
\end{abstract}


\section{Introduction}

For some time our knowledge of the intrinsic properties of the Galaxy's 
population of individual planetary nebulae has been restricted by large 
uncertainties in their derived distances. \citet{zh95} suggests that the 
{\it average} uncertainty in the distances cited to Galactic planetary 
nebulae is in the range 35-50\%. Others are less optimistic. Such a large 
scatter may not be surprising, given that planetary nebulae exhibit various 
morphologies and span a large range in mass \citep{kw05}.

In contrast, well-studied open clusters have distances and reddenings that 
are established to much greater precision, with distance uncertainties as 
small as 2.5\% being possible \citep{tb02}. Planetary nebulae established 
as members of open clusters are therefore a potential alternative means of 
calibrating their fundamental properties. With an inferred distance from 
cluster membership in conjunction with a planetary's angular diameter and 
expansion velocity, its true dimensions and age can be deduced. Cluster 
membership has the potential for a more direct calibration of the core 
mass-nebular He, C, and N abundance relationship expected in planetary 
nebulae as a result of single star evolution with asymptotic giant branch 
dredge-up \citep{ka00,cm00}. Planetary nebulae confirmed as cluster members 
would enhance their importance as calibrators for the Shklovsky relation 
\citep{of06} or other similar methods used to establish their distances 
\citep{bl01}. On a cautionary note, significant improvement in such 
relationships may not be possible if the observed scatter is intrinsic.

Several factors conspire to reduce the probability of observing a planetary 
nebula associated with an open cluster. First, the effective sample of 
planetary nebulae includes a large number of objects that appear to populate 
the Galactic bulge \citep{as71,zi75}, according to catalogue statistics 
\citep{ko01} on their distribution along the Galactic plane (Figure 1), as 
well as their observed radial velocities \citep{of06}. Potential calibrators 
lying in nearby open clusters are greatly reduced in number when that 
population is excluded, although many spatial coincidences still exist 
\citep{zi75}. Associated open clusters with ages of less than 
$\sim28\times 10^6$ years ($\log(\tau)\leq7.5$) are likely to be excluded, 
since stellar evolutionary models indicate that the end products of their 
evolved components are Type II supernovae explosions. Current knowledge of 
stellar evolution suggests that the immediate precursors of C/O white 
dwarfs were planetary nebulae central stars that did not undergo core 
carbon ignition.

In addition to a small sample size, the detection of an association between 
a planetary nebula and an open cluster is further hampered by the short 
lifetimes of planetary nebulae. Models indicate that their main-sequence 
progenitors were stars of $1-6.5\;M_{\sun}$ \citep[e.g.,][]{we00}, with an 
upper limit of $\sim8\;M_{\sun}$ being possible for production of Ne white 
dwarfs. The lifetime of the planetary nebula stage is very sensitive to 
initial progenitor mass and subsequent mass loss \citep[e.g.,][]{sb96}, and 
varies significantly for main-sequence turnoff ages greater than 
$\sim28\times 10^6$ years, with estimates ranging from $10^3$ to $10^5$ 
years \citep{sb96,ka00}. The most common age for nearby Galactic open 
clusters is $\sim100\times 10^6$ years ($\log(\tau) \simeq 8$), according 
to the catalogue compilation of \citet{di02} summarized in Figure 2. That 
corresponds to a main-sequence turnoff mass of $M_{TO}\simeq 4\;M_{\sun}$. 
The lifetime of planetary nebulae associated with such progenitors is of 
order $10^3$ years, essentially instantaneous on the Galactic stage.

It is of interest to note that many planetary nebulae with massive central 
stars are found in the field, which is populated by the remnants of 
dissolved open clusters. Such clusters exceed the number of bound open 
clusters by a sizeable order \citep{la03}, which suggests that, despite the 
short lifetimes of planetary nebulae with massive central stars, increasing 
the statistical sampling of possible spatial coincidences between 
planetaries and clusters may lead to successful detections. The usefulness 
of such surveys at extragalactic scales by \citet{lr06} and \citet{ma06} 
is therefore obvious: larger statistics dominate and planetary nebulae are 
readily discernable, as demonstrated by their success as standard candles 
\citep{ja89}. The success of the Macquarie/AAO/Strasbourg H$\alpha$ (MASH) 
survey \citep{pa06} in detecting large numbers of additional Galactic 
planetary nebulae has also been extremely useful in revealing additional 
coincidences with Galactic clusters.

The discovery of planetary nebulae within globular clusters \citep{ja97} 
raises a pertinent point that must be considered. If we consider 
$1-1.5\;M_{\sun}$ as a strict lower mass limit for the progenitors of 
planetary nebulae \citep{kw05,of06}, then, for the ages assigned to globular 
clusters, corresponding to main sequence turnoffs of less than $1M_{\sun}$, 
one must invoke binarity (mass transfer) to resolve the resulting 
discrepancy. That supports the scenario of \citet{dm06}, \citet{so06}, 
and \citet{zi07}, who argue that a large fraction of observable planetaries 
may indeed stem from binary systems. Consequently, if a planetary 
nebula/open cluster association is established, we must be aware of the 
possibility that binarity might negate possible predictions for progenitor 
mass on the basis of the cluster's implied age from its main sequence 
turnoff.

In this paper we consider the possibility of cluster membership for a number 
of planetary nebulae that are located in close proximity to open clusters 
lying in their lines of sight. The often cited cases for planetary 
nebula/open cluster associations include the cluster and nebula designated 
as NGC 2818, as well as A9 in NGC 1912 (M38) and NGC 2438 in NGC 2437 (M46), 
but lesser known cases are also considered.

\section{Suspected PN/Open Cluster Associations}

Lubos Kohoutek has compiled an on-line list of suspected planetary 
nebula/open cluster associations, somewhat different from that given by 
\citet{zi75}, that is reproduced in Table 1. To that list we have appended 
a new case involving the planetary nebula M 1-80 and the open cluster 
Berkeley 57, and also consider two planetaries that may be outlying members 
of IC 2488 \citep{pe87}.

Table 1 includes, where available, a planetary nebula identifier tied to 
its Galactic co-ordinates, along with information on the dimensions of the 
spatially adjacent cluster and the angular separation of the planetary from 
the cluster center. Open clusters are generally larger than the obvious 
concentrations of stars comprising their core regions \citep{kh69,ni02}, so 
we list in columns 4 and 5 estimates for the nuclear radius, $r_n$, and 
coronal radius, $R_C$, of each cluster, where the two dimensions are tied to 
the definitions of \citet{kh69} based upon linear star counts. The cluster 
nucleus comprises the dense central region of a star cluster that is obvious 
to the eye, whereas the cluster corona is the much lower density outer region. 
Information on both parameters is not generally available for most clusters, 
although estimates of cluster angular diameter given by \citet{di02} closely 
match the diameters of cluster nuclei defined by \citet{kh69}. Coronal radii 
can be 2.5 to 10 times larger \citep{kh69}, and the best means of estimating 
that parameter seems to be by scaling the large angular diameters for open 
clusters cited by \citet{ba50}, as described in the table footnotes. The 
last values are crude approximations at best, but at least provide a sense 
of scale for establishing if a planetary nebula's angular separation from a 
cluster is consistent with a {\it bona fide} spatial coincidence.

Table 2 outlines the qualitative framework used to determine if the suspected 
associations are co-spatial. The primary criteria are the differences in 
radial velocity and color excess between the planetary nebula and cluster 
($\Delta V_R$, $\Delta E_{B-V}$), and the ratio of the estimated distances 
($D_R$). The following parameters are also considered: the apparent size of 
the objects, their angular separation, and Galactic location. Because of the 
large number of planetary nebulae lying in the direction of the Galactic 
bulge noted earlier, there is a natural bias towards purely line of sight 
coincidences with open clusters for planetaries lying in that direction 
(recall Figure 1).

The interstellar reddenings of planetary nebulae cited throughout this study 
were usually derived from the standard constant of extinction, $c$, via the 
following generic approximation:
\begin{displaymath}
E_{B-V}\simeq0.77\times c 
\end{displaymath}
from \citet{of06}, where $c$ is related to the logarithmic extinction at 
H$\beta$. The resulting reddenings may be systematically higher than 
those for stars in the surrounding fields if there is inherent 
self-absorption by dust within the planetary nebulae themselves.

Reddening is not necessarily a strong criterion for a spatial coincidence. 
The spatial distribution of interstellar extinction near the Sun \citep{nk80} 
is clearly defined, and indicates that the dust is concentrated in distinct 
clouds rather than more or less uniformly distributed along the Galactic 
plane. Between the dust clouds along some lines of sight are large gaps, of 
a kiloparsec or more, within which all stars share similar reddenings. Small 
spatial variations in reddening can be attributed to density variations 
within the clouds themselves.

\subsection{M 3-20 and Trumpler 31 ($\ell\simeq2\degr$)}

The open cluster Trumpler 31 was studied photographically on the {\it RGU} 
system by \citet{sv66}. \citet{ja82} obtained a cluster distance of $d=1.86$ 
kpc and a reddening of $E_{B-V}=0.43$ after transforming the data to the 
{\it UBV} system. The cluster is not an obvious concentration of stars on 
POSS images of the field, and star counts are needed to assess its reality. 
A color excess of $E_{B-V}\simeq1.10\pm0.08$ derived for the planetary 
nebula M 3-20 \citep{ty92} places it beyond $\sim2$ kpc according to the 
extinction maps of \citet{nk80}. The estimated distance is $d\simeq5000\pm350$ 
pc \citep{zh95}, which, in conjunction with the reddening, small angular 
size, and Galactic location towards the Galactic bulge (Figure 2), confirm 
the planetary nebula as a background object to the cluster Trumpler 31, which 
may not exist.

\subsection{M 1-80 and Berkeley 57 ($\ell\simeq108\degr$)}

Berkeley 57 is an older cluster recently examined by \citet{ha04}, who 
derived a distance of $d=4150$ pc, a reddening of $E_{B-V}=0.75$, and an 
age of $14\times 10^8$ years ($\log(\tau)=9.14$). The planetary nebula 
M 1-80 is located $10\arcmin$ from the cluster, and is estimated to lie 
at a distance of $d=5250\pm500$ pc \citep{zh95} with a reddening of 
$E_{B-V}\simeq0.54\pm0.11$ \citep{ty92}. Reddening alone constrains the 
distance of both objects only to somewhere in the interval $2-6$ kpc 
\citep{nk80}, so the case rests mainly on the similarity of the distance 
estimates.

Star counts compiled from 2MASS data (Figure 3) lead to an estimated 
cluster nuclear radius of $r_n=5\arcmin$ \citep[see][]{kh69}, which means 
that the planetary lies only $2\times r_n$ from the cluster center. It is 
therefore a potential member of the cluster corona, adding further 
interest to the study of its potential association with Berkeley 57. The 
next step would be to use spectroscopic observations of the many 
evolved cluster giants to establish the cluster's radial velocity, for 
comparison with the value of $V_{R}=-58\pm10$ km s$^{-1}$ derived for 
M 1-80 by \citet{du98}.

\subsection{A9 and NGC 1912 (M38) ($\ell\simeq172\degr$)}

Various literature studies of the parameters for NGC 1912 (M38) generated a 
wide range of distance estimates for the open cluster, the likely reason 
being the 0.4-mag. spread in color excesses for member stars. The cluster 
distance is $d\simeq970\pm40$ pc \citep{tu76b} when that is taken into 
account. An independent distance estimate was obtained using data from the 
{\it Two Micron All Sky Survey} \citep[2MASS,][]{cu03} to construct a 
{\it J} versus {\it J--H} color-magnitude diagram, shown in Figure 4. 
Isochrones tailored specifically to the 2MASS system were obtained from the 
{\it Padova Database of Stellar Evolutionary Tracks and Isochrones} 
\citep{bo04}. A sub-solar metallicity solution ($Z=0.008$) provided the 
best visual fit, and is supported by the work of \cite{jh75}, who determined 
a cluster metallicity of [Fe/H]$\simeq-0.35$, which corresponds to 
$Z\simeq0.009$ according to the relationship found by \citet{be94}. The 
following relationships were adopted between extinction and color excess 
in the infrared and optical regions: $A_J=0.276\times A_V$, 
$E_{J-H}=0.33\times E_{B-V}$ \citep{bb05,du02}. The canonical distance 
modulus relation was reformulated and evaluated as: 
\begin{displaymath}
\log(d)=0.2[J-M_{J}-0.84(E_{J-H}\times R_V)+5]
\end{displaymath}
The results, displayed in Figure 4, yield a distance of $d=1050\pm150$ pc, 
a reddening of $E_{B-V}=0.27\pm0.03$, and an age of $18\times 10^7$ years 
($\log(\tau)=8.25\pm0.15$), essentially confirming previous estimates of a 
low reddening and a distance near 1 kpc.

With regard to the planetary nebula A9, a distance degeneracy has emerged, 
with both nearby, $d\simeq4000$ pc, $E_{B-V}\simeq1.05$ \citep{ka90}, and 
$d=5050$ pc \citep{ph04}, and distant, $d=8900\pm6100$ pc \citep{zh95}, 
solutions being advocated. The planetary's large apparent diameter of 
$30\arcsec$, measured using the {\it Aladin} environment \citep{bo00}, would 
seem to favor the nearer estimates. The extreme faintness of the central 
star \citep{kw88} and the large reddening of the planetary nebula, which 
implies a distance in excess of $\sim4$ kpc \citep{nk80} for A9, almost 
certainly place the planetary at a much greater distance than the cluster 
NGC 1912. A radial velocity of $V_{R}=-1.0\pm0.6$ km s$^{-1}$ is available 
for a red giant member of the cluster \citep{gl91}, but the radial velocity 
of A9 has not yet been measured. Presumably it would merely serve to 
confirm that the two are unrelated.

\subsection{NGC 2438 and NGC 2437 (M46) ($\ell\simeq232\degr$)}

The location of the planetary nebula NGC 2438 relative to the open cluster 
NGC 2437 (M46) is visually supportive of an association, given the planetary 
nebula's breadth, brightness, and proximity to the cluster core (Figure 5). 
Three estimates for the distance \citep{zh95} and reddening \citep{ty92} to 
the planetary nebula yield mean values of $d\simeq1775\pm630$ pc and 
$E_{B-V}\simeq0.17\pm0.08$. Both are in general agreement with a zero-age 
main sequence (ZAMS) and isochrone fit to 2MASS photometry for M46 (Figure 
6) that yields values of $d=1700\pm250$ pc, $E_{B-V}=0.13\pm0.05$, and an 
age of $22\times 10^7$ years ($\log(\tau)=8.35$). Color excesses increase 
along this line of sight from $\sim0.1$ to $\sim0.3$ at distances beyond 
$\sim1.5$ kpc \citep{nk80}, which confirms the distances estimated for both 
the cluster and the planetary nebula. The color excesses for both are also 
similar enough to confirm that they share the same space reddening. The 
case for a physical association therefore rests upon their space motions.

Early studies of the radial velocity of NGC 2438 and M46 by \citet[][citing 
measures by Struve]{cu41} and \citet{od63} indicated a difference of $\Delta 
V_R\simeq30$ km s$^{-1}$ between the objects, which suggests that the pair 
constitutes a spatial coincidence only. A cluster red giant spectroscopic 
binary has an identical systemic velocity \citep{me89} as that obtained for 
cluster dwarf members by Cuffey (see Table 3). However, \cite{pk96} 
rekindled interest in a possible planetary nebula/open cluster association 
when they found similar radial velocities for both. While it is conceivable 
that the early radial velocity measures for cluster stars, which were made 
from spectrograms obtained from the northern hemisphere, might be affected 
by spectrograph flexure or by the presence of spectroscopic binaries in the 
sample, it is noteworthy that the radial velocity measured for the planetary 
nebula is similar to more recent measures (Table 3). If the radial velocity 
measurements of \citet{pk96} are reliable, then there is a good case for a 
physical association of NGC 2438 with M46. But additional velocity measures 
are clearly needed to strengthen the case, given that proper motions may not 
provide a suitable test in this instance \citep{od63}.

\subsection{NGC 2452 and NGC 2453 ($\ell\simeq243\degr$)}

The derived distances to the open cluster NGC 2453 are unsatisfactorily 
varied, as Table 4 summarizes. Field star contamination may be important in 
this case, given that the cluster main sequence is dominated by B-type 
stars, which also populate the Puppis OB associations, and an extension of 
the Perseus arm behind them \citep{pj81}, which lie along the same line of 
sight. The distances cited from two deep CCD studies by \citet{ma95} 
and \citet{mo01} are favored because the main sequence morphology is well 
defined. The latter study of NGC 2453 implies a distance of $d=5250$ pc, 
a reddening of $E_{B-V}=0.50$, and an age of $40\times 10^6$ years 
($\log(\tau)=7.6$). The parameters for the planetary nebula NGC 2452 
generally agree with those of the cluster, although the cited distance 
of $d\simeq2950\pm420$ pc \citep{zh95} and reddening of 
$E_{B-V}\simeq0.36\pm0.12$ \citep{ty92} might suggest that the planetary 
nebula lies foreground to the cluster. The reddening along this line of 
sight remains unchanged at $E_{B-V}\simeq0.6$ for distances in excess 
of $\sim2$ kpc \citep{nk80}, so the small difference in color excesses 
is not useful for distance discrimination.

With reference to available radial velocities, \citet{mf74} obtained a 
value of $V_{R}=67\pm14$ km/s for a cluster B5 star ideally positioned as 
an evolved main-sequence member in the cluster color-magnitude diagram. 
Despite the large uncertainty in the velocity and the fact that the 
spectrogram displayed double lines, the value is very similar to measures 
for the planetary nebula: $V_R=62.0\pm2.8$ km s$^{-1}$ \citep{me88}, and 
$V_R=65\pm3$ km s$^{-1}$ \citet{du98}. Additional radial velocity 
measurements for established cluster members are needed to assess the 
viability of the case further, although existing data do not rule out a 
possible spatial coincidence.

It is of interest to note that \citet{cm00} concluded that NGC 2452 was 
among the most massive planetary nebulae in their sample of $\sim100$. 
Their argument was based on the abundance ratio N/O, which is a tracer of 
mass for the progenitor star via the dredge-up scenario. Coincidentally, 
the cluster's young age also implies a massive progenitor of 
$M_{TO}\simeq6.5\;M_{\sun}$ (see Figure 2).

\subsection{NGC 2818: Planetary Nebula and Cluster ($\ell\simeq262\degr$)}

The well known spatial coincidence of the planetary nebula NGC 2818 with 
its surrounding cluster is an example of a case that visually supports an 
association (Figure 7). \citet{pe89} determined a distance of $d=2300$ pc 
and a reddening of $E_{B-V}=0.18$ for the cluster, consistent with the 
parameters derived for the planetary nebula: $d=2660\pm830$ pc \citep{zh95}, 
and $E_{B-V}\simeq0.28\pm0.15$ \citep{ty92}. Equally encouraging are radial 
velocities from low dispersion spectra (230 {\AA} mm$^{-1}$) by \citet{ti72} 
for two A-type stars in the cluster that yielded $V_{R}=3\pm20$ km s$^{-1}$, 
compared with $V_{R}=8\pm13$ km s$^{-1}$ obtained for the planetary nebula. 
Such evidence, in conjunction with the general agreement in distance and 
reddening, has been the basis for the conclusion that the two are associated.

More recent results suggest otherwise. A comprehensive radial velocity study 
of stars in the cluster field by \citet{me01} yields a cluster radial 
velocity from 15 red giant members of $V_{R}=20.7\pm0.3$ km s$^{-1}$, while 
the radial velocity of the planetary nebula is established to be 
$V_{R}=-0.9\pm2.9$ km s$^{-1}$ \citep{du98} and $V_{R}=-1\pm3$ km s$^{-1}$ 
\citep{me88}, consistently smaller than the velocity of the cluster. The 
greater precision of recent estimates results in a velocity discrepancy of 
$\Delta V_{R}=22$ km s$^{-1}$ \citep{me01}, implying a spatial coincidence 
rather than a physical association, as concluded by \citet{me01}.

\subsection{VBRC 2 \& NGC 2899 and IC 2488 ($\ell\simeq277\degr$)}

\citet{pe97} conducted an extensive study of the planetary nebula VBRC 2 
and derived a distance of $d=1200\pm200$ pc and a reddening of $E_{B-V}=0.38$. 
The values are consistent with the parameters found for the cluster IC 2488 
by \citet{cl03}, who derived a distance of $d=1250\pm120$ pc, a reddening of 
$E_{B-V}=0.24\pm0.04$, and a radial velocity of $V_{R}=-2.63\pm0.06$ km 
s$^{-1}$. Those values are smaller than the estimates of $d=1445\pm120$ pc 
and $E_{B-V}=0.26\pm0.02$ obtained for IC 2488 by \citet{pe87}.

There may be a tendency to dismiss an association between the cluster and 
planetary because of their large apparent separation ($\simeq54\arcmin$), 
despite the consistent correlation among the parameters. Star counts of the 
field were therefore made using data available from the 2MASS survey 
(Figure 8). The data highlight IC 2488's broad extent 
($r_n\simeq17\arcmin-18\arcmin$) and indicate that the planetary nebula 
VBRC 2 lies within $\simeq 3-4$ cluster nuclear radii. Since cluster 
coronae typically extend anywhere from 2.5 to 10 times beyond their nuclear 
radii \citep{kh69}, and the coronae of star clusters in the outer Galaxy 
are larger on average than those in the inner regions \citep{ni02}, it may 
be premature to dismiss a possible association based solely on arguments 
of separation. It is of interest to note that a large fraction of 
short-period Cepheids, potential progenitors of planetary nebulae, fall 
within the coronae of their constituent clusters \citep{tu85}. A final 
decision on the case must therefore await a radial velocity for the 
planetary nebula, to assess its potential as a cluster member properly.

Published estimates for the parameters of the planetary nebula NGC 2899 
imply a distance of $d\simeq1560\pm570$ pc and a reddening of 
$E_{B-V}\simeq0.32\pm0.24$ \citep{zh95,ty92}, consistent with the parameters 
for IC 2488. \citet{du98} measured the radial velocity of the planetary 
nebula to be $V_{R}=3.4\pm2.8$ km s$^{-1}$, which differs slightly but only 
by slightly more than 2$\sigma$ from the cluster value. NGC 2899 is in the 
same situation as VBRC 2, since it also lies nearly as far from the cluster 
center, yet possibly within the corona. Remeasuring the color excess and 
radial velocity of the planetary nebula with greater precision would help 
clarify the case for cluster membership. Thus, while not conclusive, both 
candidates offer encouraging evidence.

The reddening along this line of sight becomes larger than $E_{B-V}\simeq0.3$ 
beyond $\sim1$ kpc \citep{nk80}, so the observed color excesses for the 
cluster and planetary nebulae imply that they are reddened by foreground 
dust clouds. Reddening is therefore of little use for constraining the 
distances to the planetary nebulae. Presumably radial velocities would 
provide a stronger test for a physical association.

\subsection{ESO 177-10 and Lyng{\aa} 5 ($\ell\simeq325\degr$)}

The open cluster Lyng{\aa} 5 has not been studied since its discovery nearly 
forty-five years ago, so its parameters are essentially unknown. For this 
study we examined the field and estimated a peak in star density by eye at 
J2000.0 = 15:41:55, --56:38:38, and a corresponding nuclear radius measuring 
about $2\arcmin$. As a means of obtaining approximate values for its distance 
and reddening, data from the 2MASS survey \citep{cu03} for objects in the 
field of the putative cluster nucleus were used to construct {\it JHK} 
color-color and color-magnitude diagrams for cluster stars. The color-color 
diagram (Figure 9) suggests that there is a sizable group of reddened late 
B-type stars in the field, presumably associated with the cluster main 
sequence. The implied cluster reddening, $E_{J-H}=0.33\pm0.03$, corresponds 
to $E_{B-V}=1.18\pm0.11$. A simple ZAMS fit was used to estimate the 
distance (Figure 10), yielding $d=1950\pm350$ pc. But the values cited are 
only preliminary, and still uncertain. A few cluster stars may be bright 
enough for spectroscopic follow-up, which might confirm the derived 
parameters. Interestingly enough, the cluster main-sequence turnoff appears 
to lie roughly at B5, implying an age of $\sim50\times 10^6$ years 
($\log(\tau)=7.7$), corresponding to masses of $\geq6.5 M_{\sun}$ for 
cluster evolved components.

The cited distance of $2550\pm670$ pc \citep{zh95} to the planetary nebula 
ESO 177-10 is marginally consistent with the value obtained for the 
cluster, but the color excess of $E_{B-V}\simeq2.46\pm0.07$ for the 
planetary derived from several radio measurements \citep{ty92,ca92} 
indicates that it suffers from much heavier extinction. Such a large 
reddening implies a distance in excess of $\sim2$ kpc along this line of 
sight \citep{nk80}, lending support to the argument that the planetary 
nebula lies in the cluster background. Radial velocities would likely 
confirm that this pair represents a spatial coincidence only.

\subsection{KoRe 1 and NGC 6087 ($\ell\simeq328\degr$)}

\citet{kr89} conclude that the planetary nebula KoRe 1 is in a highly 
excited state on the basis of nearly equal spectral intensities for He II 
$\lambda4686$ and H$\beta$. According to the work of \citet{gu88}, who 
formulated a relationship between the temperature of the central star and 
the various emission line ratios, KoRe 1 is among a small percentile of 
planetaries with {\it super-high} temperature central stars ($\simeq300,000$ 
K). A correspondingly high intrinsic luminosity, faint apparent magnitude 
for the nebula and central star (near the Sky Survey limits), and a small 
apparent diameter ($14\arcsec$) for the associated planetary nebula suggest 
that it is probably much more distant than the cluster, which is nearby 
\citep[$d=902\pm10$ pc,][]{tu86}. The pair appears to represent another 
case of a spatial coincidence rather than a physical association.

\subsection{HeFa 1 and NGC 6067 ($\ell\simeq330\degr$)}

\citet{hf83} concluded that the planetary nebula HeFa 1 is probably not 
associated with the open cluster NGC 6067, on the basis of an inferred 
reddening of $E_{B-V}=0.66\pm0.04$ \citep{hf83,ty92}, which is larger 
than that of the cluster. The large color excess implies a distance not 
much greater than $\sim$1--2 kpc, according to the run of reddening with 
distance along this line of sight \citep{nk80}. The cluster reddening 
is $E_{B-V}=0.35\pm0.01$ according to \citet{wa85}, and $E_{B-V}=0.32$ 
from \citet{me93}, which places a similar constraint on its distance. 
\citet{me93} find a distance of $d=1665$ pc and an age of $17\times 10^7$ 
years ($\log(\tau)=8.22$) for the cluster, so the only difference in 
parameters between the cluster and planetary nebula is the reddening, 
which is not an ideal test of membership in this instance.

NGC 6067 is also statistically unique in that it hosts two Cepheid 
members \citep{eg83}. There is consequently an {\it a priori} probability 
of detecting a planetary nebula associated with the cluster since 
short-period Cepheids are potential progenitors of stars that produce 
planetary nebulae. A radial velocity estimate for the planetary nebula 
would help resolve the question of its possible cluster membership, since a 
cluster velocity of $V_{R}=-39.3\pm1.6$ km s$^{-1}$ has been measured 
\citep{me87}. The case for a potential physical association remains open.

\subsection{Sa 2-167 and NGC 6281 ($\ell\simeq348\degr$)}

\citet{ff74} established that the cluster NGC 6281 is nearby with a distance 
of $d=560\pm30$ pc and a reddening of $E_{B-V}=0.15\pm0.02$. The planetary 
nebula Sa 2-167, however, has a much larger color excess of $E_{B-V}\simeq2.2$, 
according to \citet{ty92}, which implies a distance in excess of $\sim2$ kpc 
\citep{nk80}. One can also note the location of the planetary towards the 
Galactic bulge. The reddening discrepancy alone indicates that the planetary 
nebula lies in the background of the open cluster NGC 6281.

\subsection{M 3-45 and Basel 5 ($\ell\simeq360\degr$)}

A reanalysis by \citet{ja82} of photographic {\it RGU} photometry for the 
cluster Basel 5 by \citet{sv66} indicates that Basel 5 is relatively nearby 
with a distance of $d=1360$ pc and a reddening of $E_{B-V}=0.39$. Conversely, 
the planetary nebula M 3-45 may be a member of the Galactic bulge population 
\citep{ma98}, given its galactic longitude and large reddening of 
$E_{B-V}\simeq1.86\pm0.01$ \citep{ty92,cu00}. As in the previous case, the
large reddening discrepancy is sufficient to indicate that the planetary 
nebula cannot be associated with the cluster.

\section{Other Possible Coincidences}

The list of potential spatial coincidences between planetary nebulae and 
open clusters presented in Table 1 is only a partial listing of the spatial 
coincidences that exist \citep[e.g.,][]{zi75}. A more exhaustive listing of 
potentially good cases is presented in Table 5, derived with the aid of 
on-line lists of planetary nebulae (and possible planetary nebulae) and 
open clusters, the latter including some rather sparse spatial groupings 
not yet confirmed as true clusters (such as the anonymous group lying near 
the planetary nebula A 69). The interesting case of AL 67$-$01 \citep{al67} 
and PHR1315-6555 was highlighted by \citet{pa06}, but there are a few other 
equally interesting coincidences, such as K4 4-41 near NGC 6846. Several 
dozen other spatial coincidences exist, but they consist mainly of planetary 
nebulae lying towards the region of the Galactic center \citep{ja04} or 
bulge in purely spatial coincidence with foreground clusters. Most are 
extremely faint planetaries of small angular diameter that are almost 
certainly background to the relatively nearby clusters. Many of the objects 
in Table 5 are deserving of further study, however.

A noteworthy feature of both Table 1 and Table 5 is that there are very 
few planetary nebulae coincident with cluster nuclei, and even those cases 
may represent cluster coronal objects seen in projection against cluster 
nuclei. A sometimes overlooked property of Milky Way open clusters is that 
their dimensions and stellar content are invariably underestimated when 
gauged on the basis of stars populating their dense nuclear regions, which 
are oversampled in CCD studies. As noted in \S2, \citet{kh69} and \citet{ni02} 
find that open clusters are surrounded by low density coronae that contain 
the bulk of their member stars. By inference, that should include the bulk 
of cluster members that evolve into planetary nebulae. Cluster coronae also 
contain a large proportion of each cluster's massive stars \citep{bu78}, 
and account for the majority of Cepheids that are associated with open 
clusters \citep{tu85}. Cluster Cepheids in the Large Magellanic Cloud 
exhibit an identical characteristic \citep{ef03}.

There is no obvious explanation for the preference of massive stars and 
Cepheids, and presumably planetary nebulae progenitors, to cluster coronae. 
A dynamical origin for coronal Cepheids was proposed by \citet{tu85}, 
in which the greater frequency of stellar encounters in dense cluster 
nuclei \citep[see ][]{tu96a} results in a higher frequency of close 
binaries and merger products there, the former, because of Roche lobe 
overflow, being less likely to produce post-main-sequence stars capable 
of reaching supergiant dimensions. The same mechanism might explain the 
apparent shortage of planetary nebulae in cluster nuclei. An alternate 
explanation in terms of a radial dependence of Jeans mass $M_J$ in 
proto-clusters was suggested by \citet{bu78} to account for the 
discrepancy with respect to massive stars.

It is of interest to speculate on the future fate of cluster members with 
masses of $M \le 8\;M_{\sun}$. Such objects eventually become Cepheids 
with pulsation periods $P \le 10$ days \citep{tu96b} or planetary nebulae 
once they pass through intermediate stages as red supergiants and 
asymptotic giant branch stars. The duration of the Cepheid phase varies 
widely from $10^4-10^5$ years for first crossings of the instability 
strip to $10^6-10^7$ years for higher crossings, an order of magnitude 
(or more) longer than the planetary nebula stage. There are $\sim30$ 
known cluster Cepheids with $P \le 10$ days \citep[e.g.,][]{tb02}, so 
statistically one might expect only a few planetary nebulae to be members 
of open clusters, and a preference for cluster coronae would seem logical. 
The survey presented here appears to confirm such expectations.

\section{Discussion}

We have yet to establish a single physical association between a planetary 
nebula and an open cluster based on a correlation between their radial 
velocities, reddenings, and distances. However, further follow-up is 
indicated for a number of cases where the evidence is suggestive, namely 
M 1-80/Berkeley 57, NGC 2438/NGC 2437, NGC 2452/NGC 2453, VBRC 2 \& 
NGC 2899/IC 2488, and HeFa 1/NGC 6067, six of the thirteen coincidences considered. Additional good cases may arise from closer examination of 
some of the other coincidences noted in Table 5, but most of the associated 
clusters are as yet unstudied, limiting further progress.

Almost all potential cluster planetary nebulae lie in cluster coronal 
regions, typically surrounding open clusters for which limited or no 
photometric data exist. The fact that very few Galactic open clusters 
have been studied to the extent that both their nuclear and coronal 
regions are examined \citep{tu96a} only compounds the situation. Further 
progress requires not only new studies of our Galaxy's many unstudied 
clusters, but studies of their coronal regions as well. Spectroscopic 
observations of potentially-associated planetary nebulae would also be 
of value.

\acknowledgements
We are indebted to the following groups for facilitating the research 
described here: the staff at Vizier, WebDA, and the NASA ADS service. We are 
particularly grateful to Andre Moitinho for sending us a copy of his CMD for 
NGC 2453, Charles Bonatto for useful discussions on taking advantage of 
data from the 2MASS survey, George Jacoby for suggestions on preparing the 
text, Gunter Cibis for drawing our attention to the possibility that some 
planetary nebulae might be cluster members, and Lubos Kohoutek for compiling 
the list of suspected PNe/OC associations that was the foundation for this 
research.

\clearpage
\begin{deluxetable}{lclccc}
\tabletypesize{\scriptsize}
\tablecaption{Possible Planetary Nebula/Open Cluster Associations. 
\label{tbl-1}}
\tablewidth{0pt}
\tablehead{
\colhead{Planetary Nebula} &\colhead{PN Identifier} &\colhead{Open Cluster} 
&\colhead{Cluster $r_n$ ($\arcmin$)\tablenotemark{a}} 
&\colhead{Estimated $R_C$ ($\arcmin$)\tablenotemark{b}} 
&\colhead{Separation ($\arcmin$)} }

\startdata
M 3-20 &G002.1$-$02.2 &Trumpler 31 &3 &24 &7 \\
M 1-80 &G107.7$-$02.2 &Berkeley 57 &5 &\nodata &10 \\
A9 &PK172$+$00.1 &NGC 1912 (M38) &10 &31 &13 \\
NGC 2438 &G231.8$+$04.1 &NGC 2437 (M46) &10 &35 &5 \\
NGC 2452 &G243.3$-$01.0 &NGC 2453 &2 &21 &9 \\
NGC 2818 &G261.9$+$08.5 &NGC 2818 &5 &24 &1 \\ 
NGC 2899 &G277.1$-$03.8 &IC 2488 &17 &\nodata &54 \\
VBRC 2 &G277.7$-$03.5 &IC 2488 &17 &\nodata &53 \\
ESO 177-10 &G324.8$-$01.1 &Lyng{\aa} 5 &5 &\nodata &2 \\
KoRe 1 &G327.7$-$05.4 &NGC 6087 &7 &31 &4 \\
HeFa 1 &G329.5$-$02.2 &NGC 6067 &7 &33 &12 \\
Sa 2-167 &G347.7$+$02.0 &NGC 6281 &4 &26 &6 \\
M 3-45 &G359.7$-$01.8 &Basel 5 &3 &\nodata &5 \\
\enddata

\tablenotetext{a}{Estimated from the angular radius cited by \citet{di02}, 
except as noted in the text.}
\tablenotetext{b}{From $R_C \simeq 35\arcmin +$ Angular Diameter \citep{ba50}.} 
\end{deluxetable}

\begin{deluxetable}{lccc}
\tabletypesize{\scriptsize}
\tablecaption{Framework for Evaluating Possible Physical Associations. 
\label{tbl-2}}
\tablewidth{0pt}
\tablehead{
\colhead{Criterion} &\colhead{Likely Member} &\colhead{Potential Member} 
&\colhead{Nonmember} }

\startdata
$\Delta V_R$ &$\le 5$ km/s &$5-10$ km s$^{-1}$ &$\ge10$ km s$^{-1}$ \\
$\Delta E_{B-V}$ &$\le 0.2$ &$0.2-0.6$ &$\ge0.6$ \\
$d{\rm (PN)}/d{\rm (cluster)}$ & $\simeq 1$ &$1-2$ &$\ge2$ \\
\enddata
\end{deluxetable}

\begin{deluxetable}{lccc}
\tabletypesize{\scriptsize}
\tablecaption{Radial Velocities for NGC 2438 and NGC 2437 (M46). 
\label{tbl-3}}
\tablewidth{0pt}
\tablehead{
\colhead{Source} &\colhead{$V_R$(PN)} &\colhead{$V_R$(Cluster)} 
&\colhead{Stars} \\
&\colhead{(km s$^{-1}$)} &\colhead{(km s$^{-1}$)} & }

\startdata
Struve \citep{cu41} &77 &$45.1\pm5.5$ &5 \\
\cite{od63} &$75\pm5$ &$48.1\pm 3.0$ &1 \\
\cite{me88} &$74\pm4$ &\nodata &\nodata \\
\cite{me89} &\nodata &$48.1\pm0.1$ &1 (orbit) \\
\cite{du98} &$75\pm2.5$ &\nodata &\nodata \\
\cite{pk96} &$60.3\pm3.6$ &$60.8\pm4.0$ &4 \\
\enddata
\end{deluxetable}

\begin{deluxetable}{lcc}
\tabletypesize{\scriptsize}
\tablecaption{Parameters for the Cluster NGC 2453. 
\label{tbl-4}}
\tablewidth{0pt}
\tablehead{
\colhead{Source} &\colhead{Distance} &\colhead{$E_{B-V}$} \\
&\colhead{(pc)} & }

\startdata
\cite{mf74} &2900 &0.47 \\
\cite{gl97} &2400 &\nodata \\
\cite{ma95} &$5900\pm200$ &\nodata \\
\cite{da99} &2400 &0.48 \\
\cite{mo06} &$5250$ &0.50 \\
\enddata
\end{deluxetable}

\begin{deluxetable}{lclccc}
\tabletypesize{\scriptsize}
\tablecaption{Additional Planetary Nebula/Open Cluster Coincidences 
($r < 15\arcmin$). 
\label{tbl-5}}
\tablewidth{0pt}
\tablehead{
\colhead{Planetary Nebula} &\colhead{PN Identifier} &\colhead{Open Cluster} 
&\colhead{Cluster $r_n$ ($\arcmin$)\tablenotemark{c}} 
&\colhead{Estimated $R_C$ ($\arcmin$)\tablenotemark{d}} 
&\colhead{Separation ($\arcmin$)} }

\startdata
NGC 6741 &G033.8$-$02.6 &Berkeley 81 &3 &\nodata &13 \\
K4 4-41 &G068.7$+$01.9 &NGC 6846 &1 &\nodata &1 \\
KLW 6 &G070.9$+$02.4 &Berkeley 49 &2 &\nodata &11 \\
K 3-57 &G072.1$+$00.1 &Berkeley 51 &1 &\nodata &12 \\
A 69 &G076.3$+$01.1 &Anon (Turner) &3 &\nodata &4 \\
Bl 2-1 &G104.1$+$01.0 &NGC 7261 &3 &22 &7 \\
FP0739-2709 &G242.3$-$02.4 &ESO 493$-$03 &4 &\nodata &8 \\
PHR0840-3801 &G258.4$+$02.3 &Ruprecht 66 &1 &\nodata &2 \\
PHR0905-5548 &G274.8$-$05.7 &ESO 165$-$09 &8 &\nodata &9 \\
Pe 2-4 &G275.5$-$01.3 &van den Bergh-Hagen 72 &1 &\nodata &9 \\
\nodata &\nodata &NGC 2910 &2 &24 &14 \\
NeVe 3-1 &G275.9$-$01.0 &NGC 2925 &5 &26 &12 \\
Hf 4 &G283.9$-$01.8 &van den Bergh-Hagen 91 &3 &\nodata &14 \\
He 2-86 &G300.7$-$02.0 &NGC 4463 &2 &22 &3 \\
PHR1315-6555 &G305.3$-$03.1 &AL 67$-$01 &2 &\nodata &1 \\
PHR1429-6043 &G314.6$-$00.1 &NGC 5617 &5 &25 &1 \\
vBe 3 &G326.1$-$01.9 &NGC 5999 &2 &25 &5 \\
\enddata
\tablenotetext{c}{Estimated from the angular radius cited by \citet{di02}, 
except as noted in the text.}
\tablenotetext{d}{From $R_C \simeq 35\arcmin +$ Angular Diameter \citep{ba50}.} 
\end{deluxetable}

\clearpage

\begin{figure}
\plotone{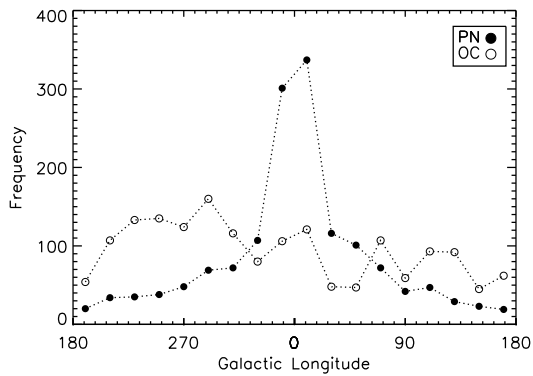}
\caption{The distribution of planetary nebulae and open clusters with 
galactic longitude, from data tabulated in the catalogs of \citet{ko01} and 
\citet{di02}. Open clusters appear to be randomly distributed along the 
Galactic plane, whereas planetary nebulae are concentrated towards the 
Galactic bulge. \label{fig1}}
\end{figure}

\begin{figure}
\plotone{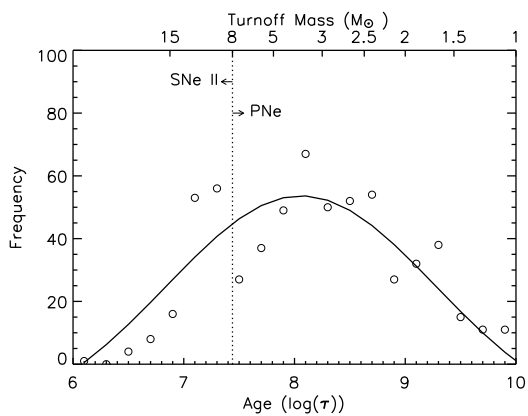}
\caption{The distribution of open cluster ages compiled from the catalog of 
\citet{di02} relative to the age/turnoff-mass relation given by \citet{ir83}. 
Open cluster ages can be described by a normal distribution with a peak near 
$\log(\tau)\simeq8$. The predicted upper turnoff mass limit, below which stars 
may evolve to produce planetary nebulae, is $M\simeq8\;M_{\sun}$, above which 
Type II supernovae are expected. \label{fig2}}
\end{figure}

\begin{figure}
\plotone{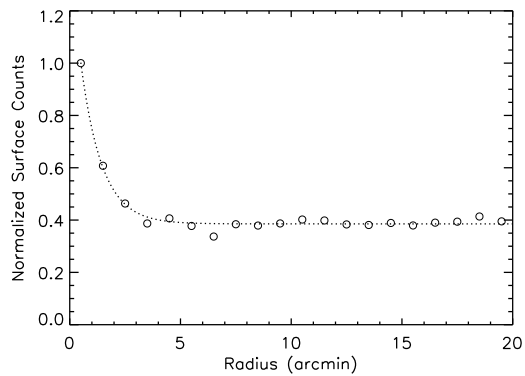}
\caption{Star counts for the open cluster Berkeley 57 derived from data in 
the 2MASS survey. \label{fig3}}
\end{figure}

\begin{figure}
\plotone{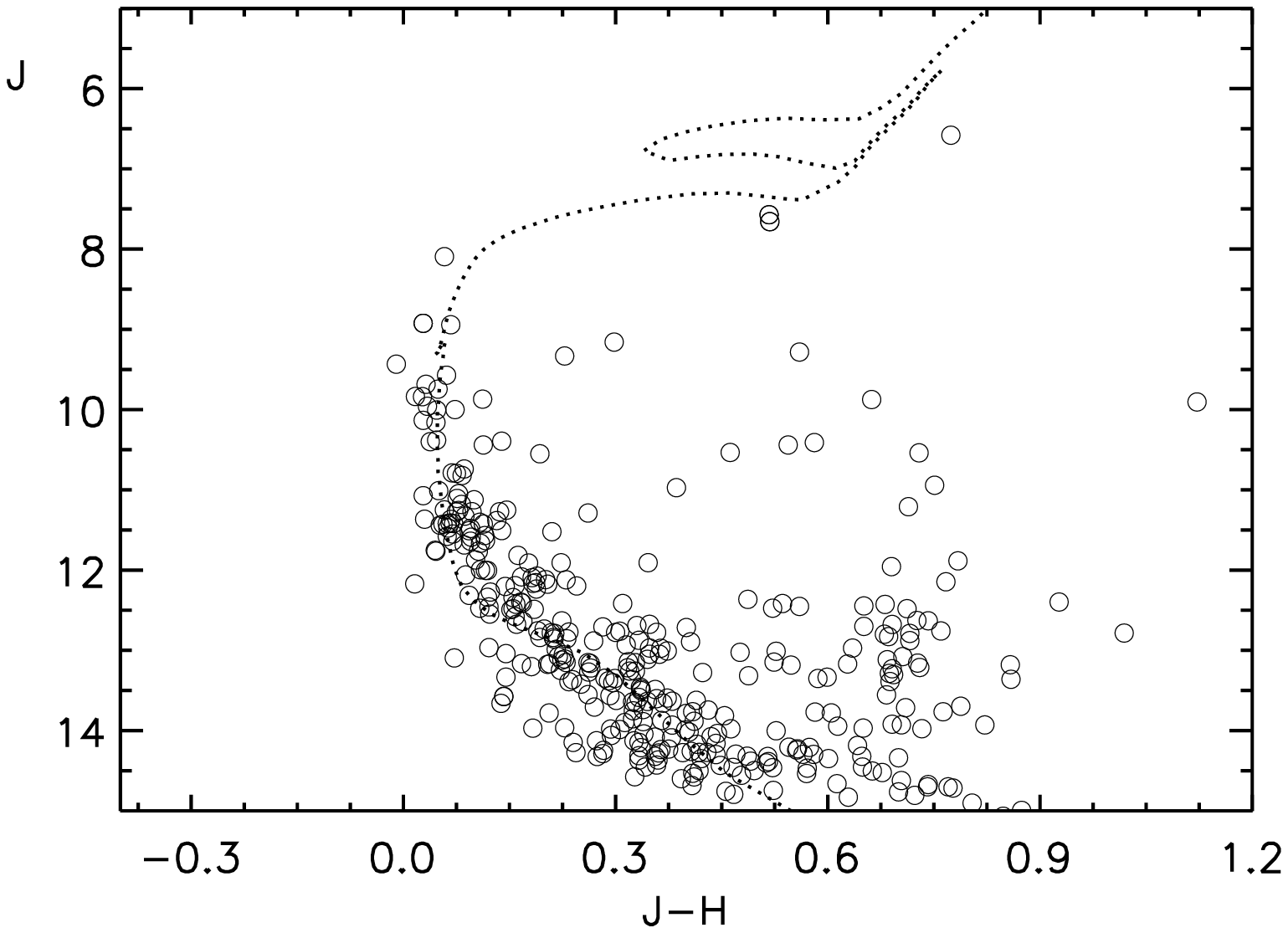}
\caption{A color-magnitude diagram for M38 (NGC 1912) constructed from 2MASS 
data. A $\log(\tau)=8.25\pm0.15$ (Z=0.008) isochrone has been fitted to the 
observations, yielding a distance of $d=1050\pm150$ pc and a reddening of 
$E_{B-V}=0.27\pm0.03$. \label{fig4}}
\end{figure}

\begin{figure}
\plotone{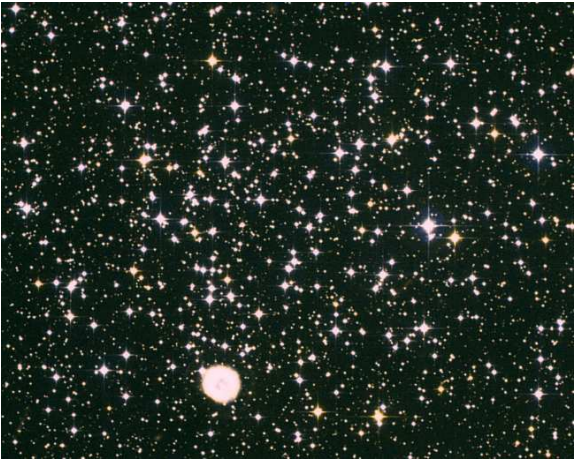}
\caption{The field of view of M46 (NGC 2437), from a combination of images 
taken at the Abbey Ridge Observatory with POSS II. \label{fig5}}
\end{figure}

\begin{figure}
\plotone{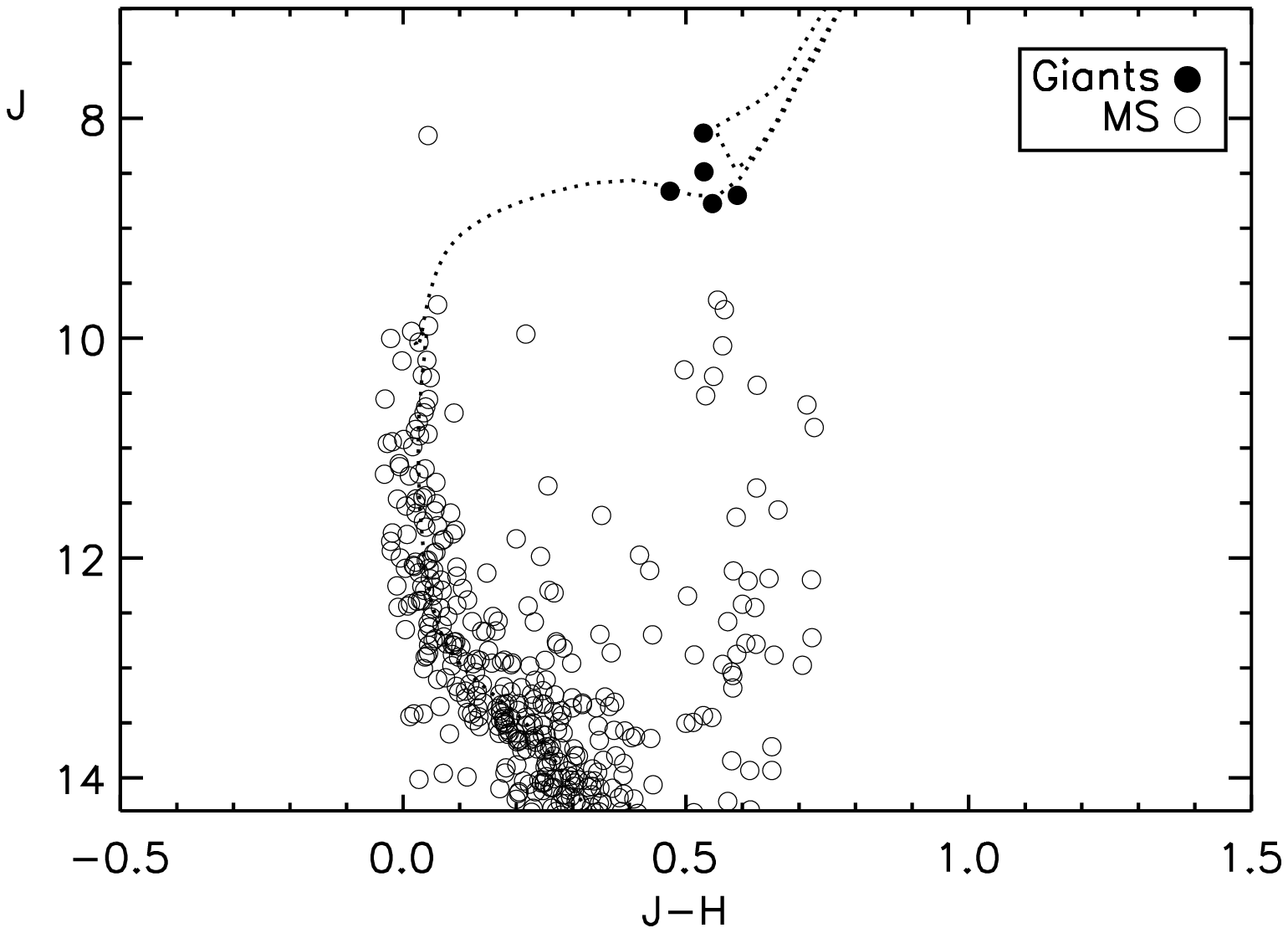}
\caption{ A color-magnitude diagram for M46 (NGC 2437) constructed from 
2MASS data. A $\log(\tau)=8.35$ isochrone has been fitted to the observations, 
yielding a distance of $d=1700\pm250$ pc and a reddening of 
$E_{B-V}=0.13\pm0.05$. \label{fig6}}
\end{figure}

\begin{figure}
\plotone{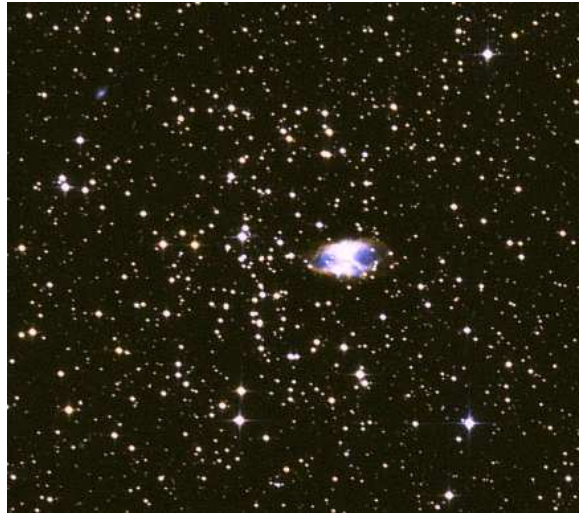}
\caption{The field of view of NGC 2818, a pseudo color image constructed from 
POSS II data. \label{fig7}}
\end{figure}

\begin{figure}
\plotone{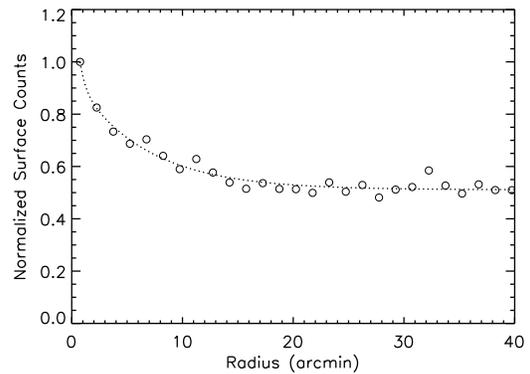}
\caption{Star counts for the open cluster IC 2488 from 2MASS data. \label{fig8}}
\end{figure}

\begin{figure}
\plotone{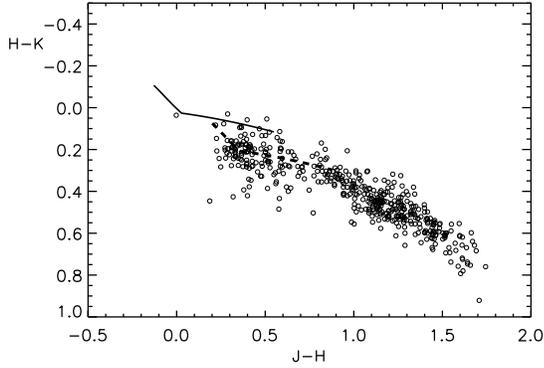}
\caption{A {\it JHK} color-color diagram for Lyng{\aa} 5 constructed from 
2MASS data. Likely cluster stars are reddened by $E_{J-H}=0.33\pm0.03$, 
which is equivalent to $E_{B-V}=1.18\pm0.11$. A reddening relation of slope 
$E_{J-H}=1.72\times E_{H-K}$ was adopted from \citet{du02,bo06}. \label{fig9}}
\end{figure}

\begin{figure}
\plotone{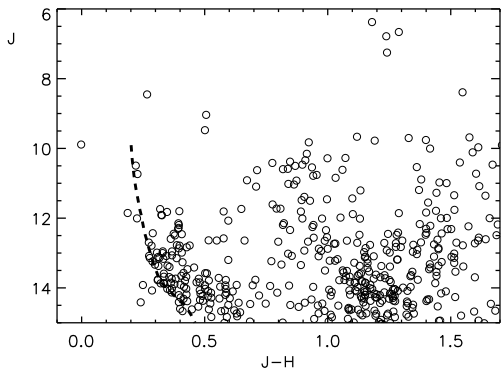}
\caption{A {\it JH} color-magnitude diagram for the open cluster Lyng{\aa} 5 
constructed from 2MASS data for stars within a $5\arcmin$ field centered 
on the J2000 co-ordinates for the cluster cited here. A ZAMS fit yields a 
distance of $d=1950\pm350$ pc for the reddening indicated in Fig. 9. 
\label{fig10}}
\end{figure}


\begin{thebibliography}{}
\bibitem[Akhundova \& Seidov(1971)]{as71} Akhundova, G.~V. \& Seidov, Z.~F.\ 
    1971, \sovast, 14, 734
\bibitem[Andrews \& Lindsay(1967)]{al67} Andrews, A.~D., \& Lindsay, E.~M.\ 
    1967, Irish Astronomical Journal, 8, 126
\bibitem[Barkhatova(1950)]{ba50} Barkhatova, K.~A.\ 1950, AZh, 27, 180
\bibitem[Bensby \& Lundstr{\"o}m(2001)]{bl01} Bensby, T., \& Lundstr{\"o}m, 
    I.\ 2001, \aap, 374, 599
\bibitem[Bertelli et al.(1994)]{be94} Bertelli, G., Bressan, A., Chiosi, 
    C., Fagotto, F., \& Nasi, E.\ 1994, \aaps, 106, 275
\bibitem[Bica \& Bonatto(2005)]{bb05} Bica, E., \& Bonatto, C.\ 2005, 
    \aap, 443, 465
\bibitem[Bonnarel et al.(2000)]{bo00} Bonnarel, F., et al.\ 2000, \aaps, 
    143, 33
\bibitem[Bonatto et al.(2004)]{bo04} Bonatto, C., Bica, E., \& Girardi, 
    L.\ 2004, \aap, 415, 571
\bibitem[Bonatto et al.(2006)]{bo06} Bonatto, C., Bica, E., Ortolani, S., 
    \& Barbuy, B.\ 2006, \aap, 453, 121
\bibitem[Burki(1978)]{bu78} Burki, G.\ 1978, \aap, 62, 159
\bibitem[Cahn et al.(1992)]{ca92} Cahn, J.~H., Kaler, J.~B., \& 
    Stanghellini, L.\ 1992, \aaps, 94, 399
\bibitem[Cazetta \& Maciel(2000)]{cm00} Cazetta, J.~O., \& Maciel, W.~J.\ 
    2000, \rmxaa, 36, 3
\bibitem[Clari{\'a} et al.(2003)]{cl03} Clari{\'a}, J.~J., Piatti, A.~E., 
    Lapasset, E., \& Mermilliod, J.-C.\ 2003, \aap, 399, 543
\bibitem[Cuffey(1941)]{cu41} Cuffey, J.\ 1941, \apj, 94, 55
\bibitem[Cuisinier et al.(2000)]{cu00} Cuisinier, F., Maciel, W.~J., 
    K{\"o}ppen, J., Acker, A., \& Stenholm, B.\ 2000, \aap, 353, 543 \bibitem[Cutri et al.(2003)]{cu03} Cutri, R.~M., et al.\ 2003, The IRSA 
    2MASS All-Sky Point Source Catalog, NASA/IPAC Infrared Science Archive
\bibitem[Dambis(1999)]{da99} Dambis, A.~K.\ 1999, Astron. Letters, 25, 10
\bibitem[De Marco(2006)]{dm06} De Marco, O.\ 2006, ArXiv Astrophysics 
    e-prints, arXiv:astro-ph/0605626
\bibitem[Dias et al.(2002)]{di02} Dias, W.~S., Alessi, B.~S., Moitinho, A., 
    \& L{\'e}pine, J.~R.~D.\ 2002, \aap, 389, 871
\bibitem[Durand et al.(1998)]{du98} Durand, S., Acker, A., \& Zijlstra, 
    A.\ 1998, \aaps, 132, 13
\bibitem[Dutra et al.(2002)]{du02} Dutra, C.~M., Santiago, B.~X., \& Bica, 
    E.\ 2002, \aap, 381, 219
\bibitem[Efremov(2003)]{ef03} Efremov, Yu.~N.  2003, Astron. Rept., 47, 1000
\bibitem[Eggen(1983)]{eg83} Eggen, O.~J.\ 1983, \aj, 88, 379
\bibitem[Feinstein \& Forte(1974)]{ff74} Feinstein, A., \& Forte, J.~C.\ 
    1974, \pasp, 86, 284
\bibitem[Gurzadian(1988)]{gu88} Gurzadian, G.~A.\ 1988, \apss, 149, 343
\bibitem[Glushkova \& Rastorguev(1991)]{gl91} Glushkova, E.~V., \& 
    Rastorguev, A.~S.\ 1991, Sov. Astron. Lett., 17, 13
\bibitem[Glushkova et al.(1997)]{gl97} Glushkova, E.~V., Zabolotskikh, 
    M.~V., Rastorguev, A.~S., Uglova, I.~M., \& Fedorova, A.~A.\ 1997, 
    Pis'ma Astron. Zh., 23, 90
\bibitem[Hasegawa et al.(2004)]{ha04} Hasegawa, T., Malasan, H.~L., 
    Kawakita, H., Obayashi, H., Kurabayashi, T., Nakai, T., Hyakkai, M., 
    \& Arimoto, N.\ 2004, \pasj, 56, 295
\bibitem[Henize \& Fairall(1983)]{hf83} Henize, K.~G., \& Fairall, A.~P.\ 
    1983, in Planetary Nebulae, IAU Symp.~103, ed. D.~R. Flower (D. Reidel 
    Publ.: Dordrecht), p. 544
\bibitem[Iben \& Renzini(1983)]{ir83} Iben, I., \& Renzini, A.\ 1983, 
    \araa, 21, 271
\bibitem[Janes \& Adler(1982)]{ja82} Janes, K.\ \& Adler, D.\ 1982, \apjs, 
    49, 425
\bibitem[Jacoby(1989)]{ja89} Jacoby, G.~H.\ 1989, \apj, 339, 39
\bibitem[Jacoby et al.(1997)]{ja97} Jacoby, G.~H., Morse, J.~A., Fullton, 
    L.~K., Kwitter, K.~B., \& Henry, R.~B.~C.\ 1997, \aj, 114, 2611
\bibitem[Jacoby \& van de Steene(2004)]{ja04} Jacoby, G.~H., \& van de 
    Steene, G.\ 2004, \aap, 419, 563
\bibitem[Jennens \& Helfer(1975)]{jh75} Jennens, P.~A., \& Helfer, H.~L.\ 
    1975, \mnras, 172, 681
\bibitem[Kaler et al.(1990)]{ka90} Kaler, J.~B., Shaw, R.~A., \& Kwitter, 
    K.~B.\ 1990, \apj, 359, 392
\bibitem[Kholopov(1969)]{kh69} Kholopv, P.~N.\ 1969, \sovast, 12, 625
\bibitem[Koester \& Reimers(1989)]{kr89} Koester, D., \& Reimers, D.\ 
    1989, \aap, 223, 326
\bibitem[Kohoutek(2001)]{ko01} Kohoutek, L.\ 2001, VizieR Online Data 
    Catalog, 4024, 0
\bibitem[K{\"o}ppen \& Acker(2000)]{ka00} K{\"o}ppen, J., \& Acker, A.\ 
    2000, in Massive Stellar Clusters, ASP Conf.~Ser., 211, ed. A. Lan\c{c}on 
    \& C. M. Boily (ASP: San Francisco), p. 151
\bibitem[Kwitter et al.(1988)]{kw88} Kwitter, K.~B., Jacoby, G.~H., \& 
    Lydon, T.~J.\ 1988, \aj, 96, 997
\bibitem[Kwok(2005)]{kw05} Kwok, S.\ 2005, J. Korean Astron. Soc., 38, 271
\bibitem[Lada \& Lada(2003)]{la03} Lada, C.~J., \& Lada, E.~A.\ 2003, \araa, 
    41, 57
\bibitem[Larsen \& Richtler(2006)]{lr06} Larsen, S.~S., \& Richtler, T.\ 
    2006, \aap, 459, 103
\bibitem[Magrini(2006)]{ma06} Magrini, L.\ 2006, in Planetary Nebulae in our 
    Galaxy and Beyond, IAU Symp. 234, ed. M.\ J. Barlow \& R.\ H. M\'{e}ndez 
    (Cambridge University Press: Cambridge), p. 9
\bibitem[Mal'kov(1998)]{ma98} Mal'kov, Y.~F.\ 1998, Astron. Reports, 42, 293
\bibitem[Mallik et al.(1995)]{ma95} Mallik, D.~C.~V., Sagar, R., \& Pati, 
    A.~K.\ 1995, \aaps, 114, 537
\bibitem[Meatheringham et al.(1988)]{me88} Meatheringham, S.~J., Wood, P.~R., 
    \& Faulkner, D.~J.\ 1988, \apj, 334, 862
\bibitem[Mermilliod et al.(1987)]{me87} Mermilliod, J.~C., Mayor, M., \& 
    Burki, G.\ 1987, \aaps, 70, 389
\bibitem[Mermilliod et al.(1989)]{me89} Mermilliod, J.-C., Mayor, M., 
    Andersen, J., Nordstrom, B., Lindgren, H., \& Duquennoy, A.\ 1989, 
    \aaps, 79, 11
\bibitem[Mermilliod et al.(2001)]{me01} Mermilliod, J.-C., Clari{\'a}, 
    J.~J., Andersen, J., Piatti, A.~E., \& Mayor, M.\ 2001, \aap, 375, 30
\bibitem[Meynet et al.(1993)]{me93} Meynet, G., Mermilliod, J.-C., \& 
    Maeder, A.\ 1993, \aaps, 98, 477
\bibitem[Moffat \& Fitzgerald(1974)]{mf74} Moffat, A.~F.~J., \& 
    Fitzgerald, M.~P.\ 1974, \aaps, 18, 19
\bibitem[Moitinho(2001)]{mo01} Moitinho, A.\ 2001, \aap, 370, 436
\bibitem[Moitinho et al.(2006)]{mo06} Moitinho, A., V{\'a}zquez, R.~A., 
    Carraro, G., Baume, G., Giorgi, E.~E., \& Lyra, W.\ 2006, \mnras, 368, L77
\bibitem[Nilakshi et al.(2002)]{ni02} Nilakshi, N., Sagar, R., Pandey, A.~K., 
    \& Mohan, V.\ 2002, \aap, 383, 153
\bibitem[Neckel \& Klare(1980)]{nk80} Neckel, Th., \& Klare, G.\ 1980, 
    \aaps, 42, 251
\bibitem[O'Dell(1963)]{od63} O'Dell, C.~R.\ 1963, \pasp, 75, 370
\bibitem[Osterbrock \& Ferland(2006)]{of06} Osterbrock, D.~E., \& Ferland, 
    G.~J.\ 2006, Astrophysics of Gaseous Nebulae and Active Galactic Nuclei 
    (University Science Books: Sausalito)
\bibitem[Parker et al.(2006)]{pa06} Parker, Q.~A., et al.\ 2006, \mnras, 
    373, 79 
\bibitem[Pauls \& Kohoutek(1996)]{pk96} Pauls, R., \& Kohoutek, L.\ 1996, 
    Astron. Nachr., 317, 413
\bibitem[Pedreros(1987)]{pe87} Pedreros, M.\ 1987, \aj, 94, 92
\bibitem[Pedreros(1989)]{pe89} Pedreros, M.\ 1989, \aj, 98, 2146
\bibitem[Pena et al.(1997)]{pe97} Pena, M., Ruiz, M.~T., Bergeron, P., 
    Torres-Peimbert, S., \& Heathcote, S.\ 1997, \aap, 317, 911
\bibitem[Peton-Jonas(1981)]{pj81} Peton-Jonas, D.\ 1981, \aaps, 45, 193
\bibitem[Phillips(2004)]{ph04} Phillips, J.~P.\ 2004, \mnras, 353, 589
\bibitem[Schoenberner \& Bloecker(1996)]{sb96} Schoenberner, D., \& 
    Bloecker, T.\ 1996, \apss, 245, 201
\bibitem[Soker(2006)]{so06} Soker, N.\ 2006, \apjl, 645, L57
\bibitem[Svolopoulos(1966)]{sv66} Svolopoulos, S.\ N.\ 1966, \zap, 64, 67
\bibitem[Tifft et al.(1972)]{ti72} Tifft, W.~G., Conolly, L.~P., \& Webb, 
    D.~F.\ 1972, \mnras, 158, 47
\bibitem[Turner(1976a)]{tu76a} Turner, D.~G.\ 1976a, \aj, 81, 97
\bibitem[Turner(1976b)]{tu76b} Turner, D.~G.\ 1976b, \aj, 81, 1125
\bibitem[Turner(1979)]{tu79} Turner, D.~G.\ 1979, \pasp, 91, 642
\bibitem[Turner(1985)]{tu85} Turner, D.~G.\ 1985, in Cepheids: Theory and 
    Observations, IAU Colloq. 82, ed. B. F. Madore (Cambridge Univ. Press: 
    Cambridge), p. 209
\bibitem[Turner(1986)]{tu86} Turner, D.~G.\ 1986, \aj, 92, 111
\bibitem[Turner(1996a)]{tu96a} Turner, D.~G.\ 1996a, in The Origins, 
    Evolution, and Destinies of Binary Stars in Clusters, ASP Conf.~Ser., 
    90, ed. E. F. Milone \& J.-C. Mermilliod (ASP: San Francisco), p. 443
\bibitem[Turner(1996b)]{tu96b} Turner, D.~G.\ 1996b, \jrasc, 90, 82
\bibitem[Turner \& Burke(2002)]{tb02} Turner, D.~G., \& Burke, J.~F.\ 2002, 
    \aj, 124, 2931 
\bibitem[Tylenda et al.(1992)]{ty92} Tylenda, R., Acker, A., Stenholm, B., 
    \& Koeppen, J.\ 1992, \aaps, 95, 337
\bibitem[Walker(1985)]{wa85} Walker, A.~R.\ 1985, \mnras, 214, 45
\bibitem[Weidemann(2000)]{we00} Weidemann, V.\ 2000, \aap, 363, 647
\bibitem[Zhang(1995)]{zh95} Zhang, C.~Y.\ 1995, \apjs, 98, 659
\bibitem[Zijlstra(2007)]{zi07} Zijlstra, A.~A.\ 2007, Baltic Astron., 16, 79
\bibitem[Ziznovsky(1975)]{zi75} Ziznovsky, J.\ 1975, Bull. Astron. Inst. 
    Czech., 26, 248
\end{thebibliography}
\end{document}